\documentclass{ws-ijmpe}
\usepackage[super]{cite}
\usepackage{graphicx}
\usepackage{lineno,hyperref}

\begin{document}

\markboth{M. Bhuyan}
{The attribute of rotational profile to the hyperon puzzle}

\catchline{}{}{}{}{}

\title{The attribute of rotational profile to the hyperon puzzle in the prediction of heaviest 
compact star}

\author{\footnotesize M. Bhuyan$^{1,2}$\footnote{Email:bunuphy@yahoo.com}, B. V. Carlson$^2$, 
S. K. Patra$^3$, Shan-Gui Zhou$^{1,4,5}$}

\address{
$^1$Key Laboratory of Theoretical Physics, Institute of Theoretical Physics, Chinese Academy 
of Sciences, Beijing 100190, China \\
$^2$Instituto Tecnol\'ogico de Aeron\'autica, 12.228-900 S\~ao Jos\'e dos Campos, S\~ao Paulo, 
Brazil \\
$^3$Institute of Physics, Sachivalaya Marg, Sainik School, Bhubaneswar 751005, India \\
$^4$Center of Theoretical Nuclear Physics, National Laboratory of Heavy Ion Accelerator, Lanzhou 
730000, China \\
$^5$Synergetic Innovation Center for Quantum Effects and Application, Hunan Normal University, 
Changsha, 410081, China }

\maketitle


\begin{abstract}
In this theoretical study, we report an investigation of the equations of state (EoSs) of 
hyper-nuclear matter and its composition as a function of density within the framework of 
effective field theory motivated relativistic mean field model. We have used G2 force 
parameter along with various hyperon-meson coupling ratios by allowing the mixing and the 
breaking of SU(6) symmetry to predict the EoSs, keeping the nucleonic coupling constant 
intact. We have estimated the properties of non-rotating and rapidly rotating configuration 
of compact stars by employing four different representative sets of equations of state. The 
obtained results of the mass and radius for the compact stars are compared with the recent 
mass observations. Further, we have studied the stability and sensitivity of rotational 
frequency (at sub-millisecond period) on the configuration of the compact stars, because the 
angular frequency is significantly smaller than the mass-shedding (Keplerian) frequency in 
slow rotation regime. Moreover, the yield of hyperon as a function of density for various 
hyperon-meson couplings are also estimated.

\end{abstract}


\section{Introduction}	

The first theoretical idea of neutron stars comes from Baade and Zwicky \cite{baade34}, being 
based on the analysis of supernova explosions. According to them, supernova explosion could be 
a transition from a star to a neutron star, consisting of closely packed neutrons in a compact 
size object. After a few years, the derivation of the full general relativistic equation of 
hydrostatic equilibrium for spherically symmetric objects, now known as 
{\it Tolman-Oppenheimer-Volkoff} (TOV) equation \cite{tolman39} came into figure. These 
equations were solved by Oppenheimer and Volkoff \cite{oppen39} assuming that the matter 
consisted of non-interacting neutrons, and found that the maximal allowed mass for this to be 
0.71$M_{\odot}$. Later it was noticed that the inclusion of nuclear energy from the interacting 
neutrons increases this value substantially \cite{chand31}. Observational proof for the 
existence of neutron star was obtained by Jocelyn Bell and Antony Hewish in 1967 by observing 
the pulsating radio beams from objects named pulsars, the rotating neutron stars with a strong 
magnetic field \cite{chand31,ghosh07,haen07}. Generally, the behavior of matter in the interior 
of a neutron star is governed by the internal pressure, where all the  neutrons obey the 
$\beta$-equilibrium condition with leptons and protons. An extended review of theoretical and 
observational aspects of neutron star, from the surface to the core, with the emphasis on their 
structure and equations of state can be found in Ref. \cite{haen07}. Further it is well known 
that the integral parameter for a foolproof structure of neutron star depends on their equation 
of state from an ideal theory, which can explain the $\beta$-equilibrium condition at high 
density. At present, a wide spectrum of different EoS for neutron star matters has been 
designed from different interactions such as Skyrme \cite{stone03}, the 
Akmal-Pandharipande-Ravenhall \cite{akmal98} and the relativistic mean field theory (RMF) 
\cite{wale04} (see Refs. \cite{shapi83,haen07,weber99} for reviews and the references therein 
for details). An extensive investigation on the EoS from these models show that they have 
almost similar properties at the nuclear saturation density (matter where the  density of 
protons is equal to that of the neutrons, $\rho_n = \rho_p$) $\rho_0\sim$ 0.16 fm$^{-3}$ but 
differ substantially at high density. In other words, the properties of nuclear matter 
predicted from different models have analogous behavior around the saturation density but 
extremely different from each other at high density under $\beta$-equilibrium. The recently 
observed massive neutron stars, J1614-2230 (with mass of 1.97$\pm$ 0.04$M_{\odot}$) 
\cite{demo10} and J0348+0432 (with mass of 2.01$\pm$ 0.04$M_{\odot}$) \cite{anto13}, have 
provided a reliable information on the existence of massive compact stars. At present, the 
uncertainties are quite large in the radius of compact stars \cite{guil11,guver10,gall12,bog13}. 
For example, using the pulse phase-resolved x-ray spectroscopy for pulsar J0437-4715 gives the 
radius of the compact star $R$ $\geq$ 11 km with 3$\sigma$ uncertainty \cite{bog13}. 
Furthermore, at high densities it is well known that there might be substantial population 
of heavy baryons (i.e. hyperons), because these become energetically favorable once the Fermi 
energy of neutrons reaches the order of their rest mass. Hence, it becomes quite necessary 
to include the hyperon matter in the study of highly dense compact objects. In the last few 
decades, a large number of systematic studies including  hyperon in the nuclear matter (i.e., 
hyper-nuclear matter or hyperon matter) were carried out {\it before} $\&$ {\it after} the 
observation of the pulsars and their identification with the neutron stars \cite{amba60,mos74,glend82,glend85,glend87,web89,kapu90,ellis91,ellis95,schaf96,hub98,sed07,bal99,bao08,schu11,long12,zhang13,qi15,lim15,biswal16,logo12,zhao12,sashi07,kras08,oert15,sharma15,haen16}. 
Most of these works demonstrate that the mass of the compact stars is reduced by 0.4$M_{\odot}$ 
where contribution of the hyperon matter is included in the EoS, challenging the knowledge of 
the perceived modern mass \cite{demo10,anto13}. Further, the inclusion of Fock channel in 
density dependent relativistic Hartree-Fock model rather effects substantially in softening 
the EoS for neutron stars \cite{long12,zhang13,qi15}. This reduction in the mass of the 
compact stars due to hyperon matter in the EoS under $\beta$-equilibrium is known as {\it 
Hyperon Puzzle} \cite{sashi07,kras08,oert15,sharma15,haen16,bom16}. The solution of this so 
called {\it Hyperon Puzzle} is not easy, it requires an additional contrivance that could 
provide a repulsion to make the EoS stiffer. At present, a few possible efforts have been 
made in this directions such as strong hyperon$-$nucleon and/or hyperon$-$hyperon interactions 
\cite{bend12,mas15,oert15,sharma15}; the inclusion of three-body forces with one or more  
hyperons \cite{yama14,lona15}; the appearance of other hadronic degrees of freedom  and phase
transition to deconfined quark matter \cite{drag14,alf07,klahn13,zdun13,alva15,benic15,bejg17} 
and considering differential rotating effects \cite{gond17}. For more detailed on the {\it 
Hyperon Puzzle}, one can follow the recent review article of Refs. \cite{chat16,laura17} and 
references therein.

It is well known that the theoretical $Mass\sim Radius$ relation depends on the rotation 
frequency and also the presence of an exotic core in massive neutron star. Within present 
scenario, all neutron stars rotate and there are many millisecond pulsars with rotation 
frequency $\geq$ 500 Hz (10 accreting X-ray pulsars and 14 radio/gamma-ray pulsars). A radio 
millisecond pulsar B1937 + 21 rotating at frequency 641 Hz \cite{back82}, remained the most 
rapidly one for more than two decades and the discovery of pulsar J1748-2446ad with a faster 
rotating of frequency 716 Hz was announced in the year 2006 \cite{hess06}. Later observations 
of X-ray burst XTE J1739-285 noticed a formation of neutron star with an initial rotation 
frequency $\nu_{\textrm in} \sim$ 1122 Hz \cite{kaar07}. On the other hand the young pulsars 
has been known to have rotational frequency $\leq$ 10 Hz and in this category (according to 
the observations so far) PSR J0537-6910 is known to be the fastest with $\nu_{\mathrm max} 
\sim$ 62 Hz \cite{mars04}. In fact, the angular velocity distribution of a neutron star 
evolves to a uniform rotation within a very short time period in any supernova event 
\cite{ande98,ande01}. Moreover, the sub-kHz frequencies are still too low to significantly 
affects the structure of massive neutron stars \cite{shap83,haen07}. However, for rapid 
rotation i.e. the sub-millisecond pulsar with super-kHz frequencies, the rotation affects 
the massive neutron stars. Reviewing the works concerning the rapidly rotating stars in 
general relativity, there are few pioneering predictions, which have been observed by 
considering some of the sophisticated aspects of the neutron star: the nuclear equation 
of state, magnetic fields, magnetic field breaking and meridional flows etc. Here we give 
some of these references for such studies: Butterworth \& Ipser (1976) \cite{butt76}; 
Komatsu {\it et al.} (1989) \cite{koma89}; Bonazzola {\it et al.} (1993) \cite{bona93}; 
Stergioulas \& Friedman (1995) \cite{fried95}; Laarakkers \& Poisson (1999) \cite{laar99}; 
Baumgarte {\it et al.} (2000) \cite{baum00}; Ansorg {\it et al.} (2002) \cite{anso02}; 
Birkl {\it et al.} (2010) \cite{birk11}; Krastev {\it et al.} (2008) \cite{kras08}. For 
more extensive global collection of literature, see Friedman \& Stergioulas (2013) 
\cite{fried13}.

Besides the calculation performed by including the $\omega$-$\rho$ cross-coupling and 
quartic terms in the RMF softens the EoS to good agreement with the recent observed mass 
of compact stars \cite{aru04,jha06,pika01,bhu13}. In addition to that the generalized model 
having the low lying octet of baryons makes it difficult to obtain a neutron star mass 
greater than 2.0$M_{\odot}$ \cite{jha06,bhu13,sharma15,haen16}. In continuation of our 
earlier work \cite{bhu13}, we here analyze the rotational attributes of neutron star with 
various hyperon-meson couplings within the framework of effective field theory motivated 
relativistic mean field model (E-RMF) \cite{aru04,furn96,sharma15,bhu13}. The degrees of 
freedom in this theory are nucleons interacting through the exchange of iso-scalar scalar 
$\sigma-$, iso-scalar vector $\omega-$, and iso-vector-vector $\rho-$ meson fields. The 
chiral effective Lagrangian (i.e. the E-RMF formalism) 
\cite{aru04,furn96,muller96,furn00,bend01,bend06,sharma07,furn97,bhu14,bhu14a,biswal16} 
is the extension of the standard relativistic mean-field (RMF) theory 
\cite{sero86,ring86,bogu77,bhu09,bhu11,rein89,ring96,meng06,hub98,lim15} with the addition 
of nonlinear scalar-vector and vector-vector self-interaction, applying the naive dimensional 
analysis and the concept of naturalness at a given level of accuracy 
\cite{aru04,furn96,muller96,sharma07,furn97,bhu14,bhu14a,biswal16}. In particular, the 
motivation behind the present work is to investigate the axisymmetric hydrostatic equilibria 
of rotating neutron star and the effects of rotational profiles in the hyperon matter under 
$\beta$-equilibrium condition at high density. In other words, this present work will 
provide a relativistic mean field descriptions of the static and rotating (sub-millisecond 
to super-millisecond) compact star properties for various hyperon-meson couplings. 

This paper is organized as follows: In Section II we discuss the theoretical setup for the 
relativistic mean field theory. The parametrization and values of their coupling constant 
are also included in this section. Section III$-$VI are assigned for the discussion of the 
results obtained from our calculation for the static and rotating compact stars. A short 
mathematical formulation for the equilibrium condition of static and rapidly rotating compact 
stars are also presented in the section IV \& V. Finally, the summary and a brief conclusion 
are given in Section VII.

\section{The relativistic mean-field theory}
The elementary theory for strong interaction to represent the complete description of nuclear 
equation of state is quantum chromodyanmics (QCD). At present, it is not conceivable to 
describe the complete picture of the hadronic matter due to its non-peturbative  properties. 
Hence, one need to endorse the perception of effective field theory (EFT) at low energy, known 
as quantum hadrodynamics (QHD) \cite{sero86,ring86,bogu77}. Now-a-days, the mean field treatment 
of QHD has been used widely to describe the properties of infinite nuclear matter 
\cite{sero86,aru04,bhu14,bhu14a} and finite nuclei \cite{bogu77,ring86,bhu09,bhu11,bhu15,hao15,meng15}.
In this theory, the nucleons are considered as Dirac particle, interact through the exchange 
of various mesons (i.e. iso-scalar scalar $\sigma$-, iso-scalar vector $\omega$-, 
iso-vector-vector $\rho$- and iso-vector-scalar $\delta$-mesons). The chiral effective 
Lagrangian (E-RMF) is proposed by Furnstahl, Serot and Tang 
\cite{furn96,mull96,furn00,bend06,sharma07,bhu14}, the extension of the standard relativistic 
mean field model \cite{sero86,ring86,bogu77,bhu09,bhu11,rein89,ring96,meng06}. In E-RMF, the 
nonlinear Lagrangian is expanded with the increasing order of the fields along with their 
derivative up to $4^{\mathrm th}$ order of interaction under naive dimensional analysis 
\cite{bend01,bend06} and the concept of naturalness \cite{mull96,furn00,sero97}. In the 
interior of a neutron star, where the density is very high, other hadronic states are produced 
\cite{sashi07,kras08,oert15,sharma15,haen16,bom16}. Thus, the considered model involves the 
full octet of baryons interacting through mesons. Finally, the truncated Lagrangian is given 
by
\begin{eqnarray}
{\cal L} & = & \sum_B \overline{\Psi}_{B}\left ( i\gamma^\mu D_{\mu} - m_{B} 
+ g_{\sigma B}{\sigma}\right){\Psi}_{B} +\frac{1}{2}{\partial_{\mu}}{\sigma}{\partial^{\mu}}
{\sigma}-\frac{1}{4} {\Omega_{\mu\nu}}{\Omega^{\mu\nu}} 
-\frac{1}{4}{R^a_{\mu\nu}}{R^{a\mu\nu}} \nonumber \\ 
&&-m_{\sigma}^2{\sigma^2}\left(\frac{1}{2} +\frac{\kappa_3}{3!}\frac{g_{\sigma N}\sigma}{m_{N}}
+\frac{\kappa_4}{4!}\frac{g^2_{\sigma N}\sigma^2}{m_{N}^2}\right) 
+\frac{1}{2}\left (1+{\eta_1}\frac{g_{\sigma N}\sigma}{m_{N}}+\frac{\eta_2}{2}
\frac{g^2_{\sigma N}\sigma^2}{m_{N}^2}\right)m_{\omega}^2{\omega_{\mu}}{\omega^{\mu}} \nonumber \\
&&+\frac{1}{2}\left(1 + \eta_{\rho}\frac{g_{\sigma N}\sigma}{m_{N}}\right)
m_{\rho}^2{\rho^a_{\mu}}{\rho^{a\mu}}+\frac{1}{4 !}{\zeta_{0}}g^2_{\omega N}
\left({\omega_{\mu}}{\omega^{\mu}}\right)^2 + \sum_l \overline{\Psi}_l 
\left(i\gamma^\mu\partial_{\mu} - m_l\right)\Psi_l.
\label{lag}
\end{eqnarray}
The subscript $B$ = n, p, $\Lambda$, $\Sigma$ and $\Xi$, denotes baryons of $m_B$, $l$ stands 
for lepton ($e^-$ \& $\mu^-$) of mass $m_l$ and $N$ for nucleons of mass $m_N$. The spin of 
all baryons is 1/2. Here, the covariant derivative $D_{\mu}$ is defined as
\begin{eqnarray}
{D_{\mu}} & = & \partial_{\mu} + ig_{\omega B}{\omega_{\mu}} + ig_{\rho B}I_{3B}{\tau^a}
{\rho^a_{\mu}},
\end{eqnarray}
where, the term $R^a_{\mu\nu}$, and $\Omega_{\mu\nu}$ are the field tensors,
\begin{equation}
R^a_{\mu\nu}= \partial_{\mu}\rho^a_{\nu} - \partial_{\nu}\rho^a_{\mu} 
+ g_{\rho N}\epsilon_{abc}\rho^b_{\mu} \rho^c_{\nu},
\end{equation}
\begin{equation}
{\Omega_{\mu\nu}} =  \partial_{\mu}\omega_{\nu} - \partial_{\nu}\omega_{\mu}.
\end{equation}
Here, $m_\sigma$, $m_\omega$ and $m_\rho$ are the masses for the baryon, $\sigma$-, $\omega$- 
and $\rho$-meson, respectively. From this Lagrangian, we derive the equation of motion and 
solve within mean field approximation self consistently. The obtained field equations for 
$\sigma$, $\omega$ and $\rho$-meson are given by
\begin{eqnarray}
m^2_\sigma \left(\sigma_0 + \frac{g_{\sigma N}\kappa_{3}}{2m_{N}}{\sigma^2_0} 
+\frac{g^2_{\sigma N}\kappa_{4}}{6m_{N}^2}{\sigma^3_0}\right)
-\frac{1}{2}m^2_{\rho}\eta_{\rho}\frac{g_{\sigma N}}{m_{N}}\rho^2_0 \nonumber \\ 
-\frac{1}{2}m^2_{\omega} \left(\eta_1\frac{g_{\sigma N}}{m_{N}} + \eta_2\frac{g^2_{\sigma N}}
{m_{N}^2}\sigma_0\right) {\omega_0^2} = \sum_{B}g_{\sigma B}\rho_{SB} 
\end{eqnarray}
\begin{eqnarray}
m^2_{\omega}\left(1 + \eta_1 \frac{g_{\sigma N} \sigma_0}{m_{N}} 
+ \frac{\eta_2}{2} \frac{g^2_{\sigma N} \sigma^2_0}{2m_{B}^2} \right){\omega}_0 
+\frac{1}{6}{\zeta_0} g^2_{\omega N}{\omega^3_0} = \sum_{B}g_{\omega B}\rho_B 
\end{eqnarray}
\begin{eqnarray}
m^2_{\rho}\left(1 + \eta_{\rho} \frac{g_{\sigma N} \sigma_0}{m_{N}}\right)\rho_{03} 
= \sum_{B}g_{\rho_B}I_{3B}\rho_B
\end{eqnarray}
For a baryon species, the scalar density, $\rho_{SB}$, and baryon density $(\rho_B)$ are 
given as,
\begin{equation}
\rho_{SB}= \frac{2J_{B}+1}{2\pi^2}\int_{0}^{k_B} \frac{M^*_B k^2 dk}{E^*_B}
\end{equation}
\begin{equation}
\rho_{B} = \frac{2J_{B}+1}{2\pi^2}\int_{0}^{k_B} {k^2 dk},
\end{equation}
where $E^{*}_{B}=\sqrt{k^{2}+{m^*}^2_{B}}$ is the effective energy and $J_{B}$ and $I_{3B}$ 
are the spin and isospin projection of baryon, respectively. The quantity $k_B$ is the Fermi 
momentum of the baryon, and $m^*=m_B-g_{\sigma B}\sigma$ is the effective mass, which can 
solve self-consistently. Now, the pressure density $\cal P$ and the energy density 
${\cal E}$ for a given baryon density are expressed as follow,
\begin{eqnarray}
\label{eqnP}
{\cal P} &=& \sum_B\frac{\gamma}{3(2\pi )^{3}}\int_{0}^{k_B}d^{3}k\frac{k^{2}}{E^{*}_{B}(k)}
+\frac{1}{ 4!}\zeta _{0}g_{\omega N}^{2}{\omega}_{0}^{4} + \frac{1}{2}\Bigg(1+\eta _{1}
\frac{g_{\sigma N} \sigma_{0}}{m_{N}}+\frac{\eta _{2}}{2}\frac{g_{\sigma N}^{2}
\sigma_{0}^{2}}{m_{N}^{2}}\Bigg)m_{\omega}^{2}{\omega}_{0}^{2}  \nonumber \\
&&\null -\Bigg(\frac{1}{2}+\frac{\kappa _{3}g_{\sigma N}\sigma _{0}}{3!m_{N}}
+\frac{\kappa _{4}g_{\sigma N}^{2}\sigma_{0}^{2}}{4!m_{N}^{2}}\Bigg) m_{\sigma}^{2}
\sigma_{0}^{2} +\frac{1}{2 }\Bigg(1+\eta_{\rho}\frac{g_{\sigma N}\sigma_{0}}{m_{N}}
\Bigg)m_{\rho }^{2}\rho_{0}^{2} + \sum_l {\cal P}_l,
\label{eqnP}
\end{eqnarray}
\begin{eqnarray}
\label{eqnE}
{\cal E} &=& \sum_B\frac{\gamma}{(2\pi )^{3}}\int_{0}^{k_B}d^{3}k E^{*}_{B}(k)
+\frac{1}{4!}\zeta _{0}g_{\omega N}^{2}{\omega}_{0}^{4} + \frac{1}{2}\Bigg(1
+\eta _{1}\frac{g_{\sigma N} \sigma_{0}}{m_{N}}+\frac{\eta _{2}}{2}\frac{g_{\sigma N}^{2}
\sigma_{0}^{2}}{m_{N}^{2}}\Bigg)m_{\omega}^{2}{\omega}_{0}^{2}  \nonumber \\
&&\null + \Bigg(\frac{1}{2}+\frac{\kappa _{3}g_{\sigma N}\sigma _{0}}{3!m_{N}}
+\frac{\kappa _{4}g_{\sigma N}^{2}\sigma_{0}^{2}}{4!m_{N}^{2}}\Bigg) 
m_{\sigma}^{2}\sigma_{0}^{2} +\frac{1}{2 }\Bigg(1+\eta_{\rho}\frac{g_{\sigma N}
\sigma_{0}}{m_{N}}\Bigg)m_{\rho }^{2}\rho_{0}^{2} + \sum_l {\cal E}_l.
\label{eqnE}
\end{eqnarray}
Here, the constant $\gamma=2$ is known as spin degeneracy parameter. The pressure density 
and energy density contribution from leptons are represented by ${\cal P}_l$ and 
${\cal E}_l$, respectively.

\section{Nuclear matter at saturation density}
\begin{figure}
\begin{center}
\includegraphics[width=1.0\columnwidth]{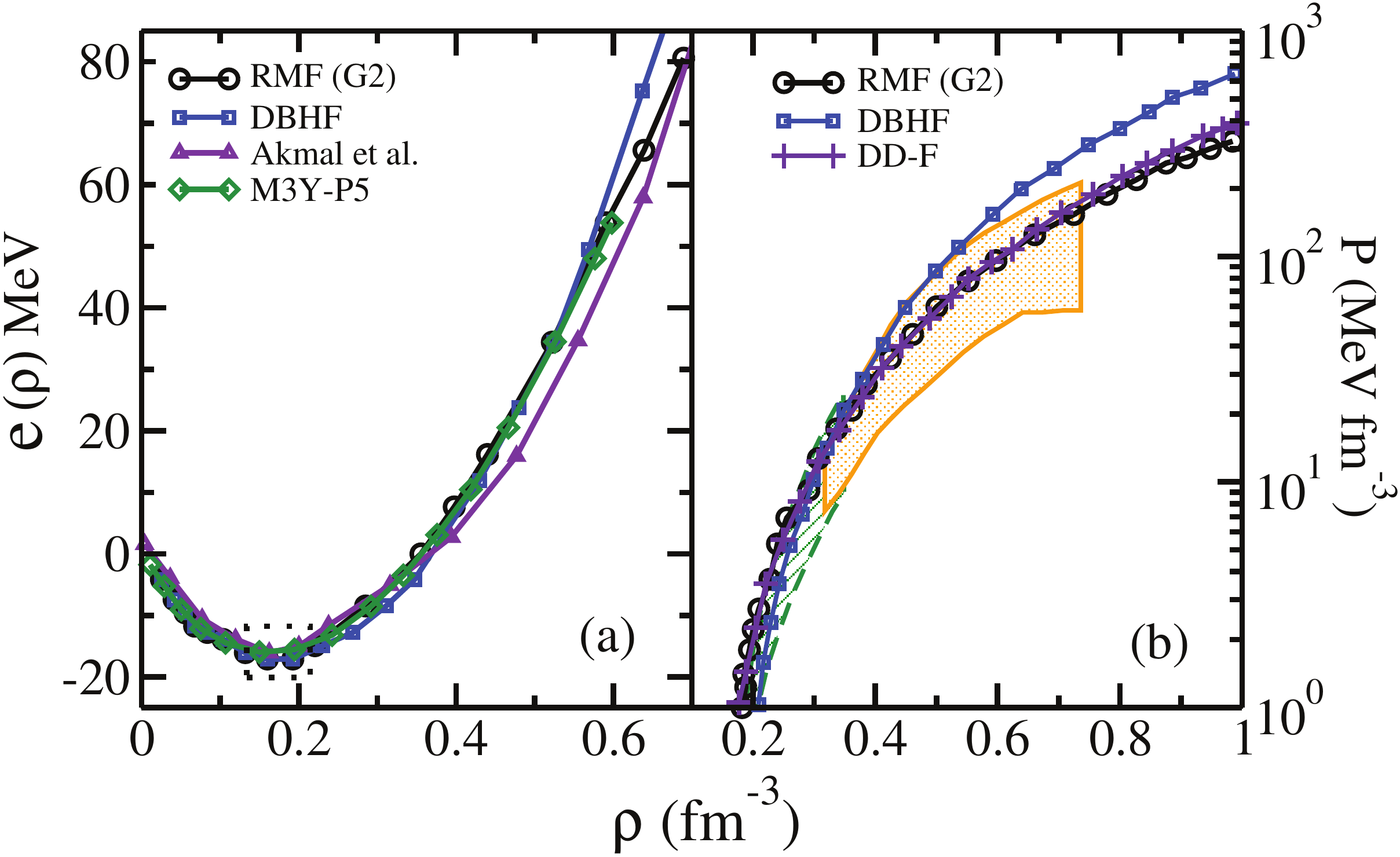}
\caption{\label{fig:1} (Color online) (a) The obtained energy per particle as a function of 
density from G2 force is compared with M3Y-P5 \cite{nakada08}, DBHF \cite{samma20}, and the 
realistic calculation \cite{akmal98}. (b) The obtained pressure density as a function of 
density from G2 force is compared with DBHF \cite{ebhf04} and DD-F \cite{klahn06}. The areas 
within solid and broken line correspond to the results extracted from HIC \cite{daneil02} and 
$K^+$ production data \cite{lynch09}, respectively. See text for details.}
\end{center}
\end{figure}

We know that the infinite nuclear matter is an essential system for the investigation of 
physical quantities relevant to heavy nuclei and compact objects like a neutron star. 
Furthermore, at saturation density, the binding energy per particle, pressure density, 
symmetry energy  and compressibility are well established physical quantities from the 
empirical and experimental observation. For general idea, the results obtained within E-RMF 
at saturation density $\rho_0$ are listed in Table 1, which are consistent with the recent 
constrained values except the slope parameter $L_{sym}$ \cite{stain14}. Following the work 
of Lattimer {\it et al.} \cite{stain14}, the values of $L_{sym}$ within G2 force is a little 
overestimated compared to their recent constraint limit. We have also shown results for the 
energy density and the pressure density as a function of baryon density using Eqs. (\ref{eqnP} 
\& \ref{eqnE}) for our calculations for G2 force in the left panel and right panel of Fig. 
\ref{fig:1}, respectively. The obtained results are compared with the M3Y-P5 \cite{nakada08}, 
DBHF \cite{samma20} and the realistic calculation done by Akmal {\it et. al.} \cite{akmal98}. 
In the right panel of Fig. \ref{fig:1}, we have also compared the results from our calculation 
with DBHF \cite{ebhf04} and DD-F \cite{klahn06} predictions. The shaded areas within solid 
and broken line correspond to the empirical results extracted from HIC \cite{daneil02} and 
$K^+$ production data \cite{lynch09}, respectively for comparison. From the figure, one can 
find that the obtained results from our calculation agree well with all others theoretical 
predictions and also with the empirical data. It is to be noted that in G2 force, the quartic 
term of the $\omega$-meson and cross coupling of the scalar field plays the major role in the 
softening of the EoS at high density with a reasonable compressibility \cite{aru04}. Being 
extension to the previous works \cite{bhu13}, the aim is to study the dependence of neutron 
star properties with an exotic core and the imprint of rotation on the mass-equatorial radius 
relation and the stability of rotating configurations.
\begin{table}[h]
\tbl{The mass of different mesons, the coupling constant, and the nuclear matter saturation 
properties for E-RMF (G2) force parameter. The masses are in MeV.}
{\begin{tabular}{@{}cccccccccccccccccccccc@{}}
\hline\hline
Meson Mass & Coupling Constant & Nuclear Matter Properties \\
\hline
$m_{\sigma}$ = 520 & $g_\sigma$ = 0.83522   & $m^*/m$ = 0.66 \\ 
$m_{\omega}$ = 782 & $g_\omega$ = 1.01560   & $\rho_0$ = 0.15 \\ 
$m_{\rho}$ = 770   & $g_\rho$ = 0.75467     & $\varepsilon (\rho_0)$ = $-$16.1 \\
                   & $\kappa_3$ = 3.2467    & $K_0$ = 214.7 \\
                   & $\kappa_4$ = 0.63152   & $E_\mathrm{sym}$ = 36.4 \\
                   & $\zeta_0$ = 2.6416     & $L_\mathrm{sym}$ = 100.7 \\
                   & $\eta_1$ = 0.64992     & \\
                   & $\eta_2$ = 0.10975     & \\
                   & $\eta_{\rho}$ = 0.3901 & \\
\hline \hline
\end{tabular}} 
\end{table}

\begin{table}[h]
\tbl{The hyperon-meson coupling ratios for baryon octet family for four different 
parametrisation. The maximum masses and the corresponding radii are obtained for 
static and stationary rotating compact star at Keplarian frequency for different 
EoSs. The quantities given within the bracket are the mass and radius of the neutron 
star without hyperons.}
{\begin{tabular}{@{}cccccccccccccccccccccc@{}}
\hline\hline
                    & Set 1 \cite{ellis91} & Set 1a \cite{ellis91} & Set 2 \cite{glend91}& 
Set 2a \cite{chia09} \\
$x_{\sigma\Lambda}$ & 0.4800               & 0.5800& 0.6104& 0.6106 \\ 
$x_{\sigma\Sigma}$  & 0.4800		   & 0.5800& 0.6104& 0.4046 \\	
$x_{\sigma\Xi}$     & 0.0000		   & 0.5800& 0.6104& 0.3195 \\
$x_{\omega\Lambda}$ & 0.5600		   & 0.6600& 0.6666& 0.6666 \\
$x_{\omega\Sigma}$  & 0.0000		   & 0.0000& 0.6666& 0.6666 \\
$x_{\omega\Xi}$     & 0.0000		   & 0.3333& 0.6666& 0.3333 \\
$x_{\rho\Lambda}$   & 0.0000		   & 0.7500& 0.6104& 1.0000 \\
$x_{\rho\Sigma}$    & 0.0000		   & 0.0000& 0.6104& 1.0000 \\
$x_{\rho\Sigma}$    & 0.0000		   & 2.0000& 0.6104& 1.0000 \\
\hline  
Static star         & 			   &       & \\
$M/M_{\odot}$       & 1.43 (2.04)   & 1.48 (2.04)   & 1.51 (2.04)   & 1.54 (2.04) \\
$R$                 & 11.02 (11.02) & 11.02 (11.02) & 11.03 (11.02) & 11.03 (11.02) \\
\hline 
Rotating star       & 			   &       &       &      \\
$M/M_{\odot}$       & 2.11 (2.43)   & 2.14 (2.43)   & 2.17 (2.43)   & 2.19 (2.43)  \\	
$R$                 & 13.35 (13.28) & 13.35 (13.28) & 13.36 (13.28) & 13.36 (13.28) \\	
\hline
\end{tabular}}
\end{table}
\section{Baryonic matter in $\beta$-equilibrium}
In the interior of neutron stars, the neutron chemical potential goes beyond the total mass 
of the proton and electron. In other words, a probabilistic neutron star composition of neutron 
star is asymmetric matter with an admixture of electrons rather than pure neutron matter. 
Further, the density is expected to be high enough ($\sim$ 7--8$\rho_0$) in the core of the 
neutron star, that undergoes a transition to other hadronic states i.e the low lying octet 
of baryons ($\Lambda^0$, $\Sigma^0$, $\Sigma^+$, $\Sigma^-$, $\Xi^0$, and $\Xi^-$) apart from 
the neucleons are produced \cite{glend85,bend01,glend97,schaf99}. For stars, the strongly 
interacting particles are baryons, the composition is determined by the requirements of 
charge neutrality and $\beta-$equilibrium conditions under the weak processes 
\cite{sharma07,bod91,weiss12}, $B_1 \to B_2 + l + {\overline \nu}_l$ and $B_2 + l \to B_1 + 
\nu_l$. Here, $B_1$ \& $B_2$, $l$, $\nu$ and $\overline{\nu}$ are the baryons, leptons, neutrino 
and anti-neutrino respectively. After deleptonization, the charge neutrality condition yields, 
\begin{equation}
q_{tot} = \sum_B q_B k_B^3/(3\pi^2) + \sum_{l=e,\mu} q_l k_l^3/(3\pi^2) = 0.
\label{eqnC}
\end{equation}
Here, the $q_B$ and $q_l$ correspond to the electric charge of baryon ({\it B}) and lepton 
species ({\it l}), respectively. Since the time scale of a star is effectively infinite compared 
to the weak interaction time scale, the strangeness quantum number is therefore not conserved 
in a star. The net strangeness is determined by the condition of $\beta-$equilibrium, which 
for baryon is then given by $\mu_B = b_B \mu_n - q_B \mu_e$, where $\mu_B$ is the chemical 
potential of baryon number $b_B$. Thus the chemical potential of any baryon can be obtained 
from the two independent chemical potentials $\mu_n$ and $\mu_e$ of neutrons and electrons, 
respectively. Hence, the equilibrium composition of the star is obtained by solving the set 
of equations for the chemical potential in conjunction with the charge neutrality condition 
in Eq. \ref{eqnC} at a given baryon density. On the basis of the quark model, one can assume 
that the hyperons interact with the mesons in two distinct modes by permitting the mixing and 
breaking of SU(6) symmetry, keeping nuclear coupling constant intact 
\cite{ellis91,schaf94,bog81,glend91,lu03,chia09,lu14}. We have used these two methods of 
parametrisations: (i) same coupling ratios as assumed by the quark model 
\cite{bog81,ellis91,schaf94} (see Set 1 \& 1a of Table 2) and (ii) different couplings 
strength for different baryons \cite{glend91,chia09,lu14} (see Set 2 \& 2a of Table 2) to 
deals with the octet of baryons in the EoS of compact star. The adopted hyperon-meson coupling 
ratios  and their values for the present studies are listed in Table 2.

\begin{figure}
\begin{center}
\includegraphics[width=1.0\columnwidth]{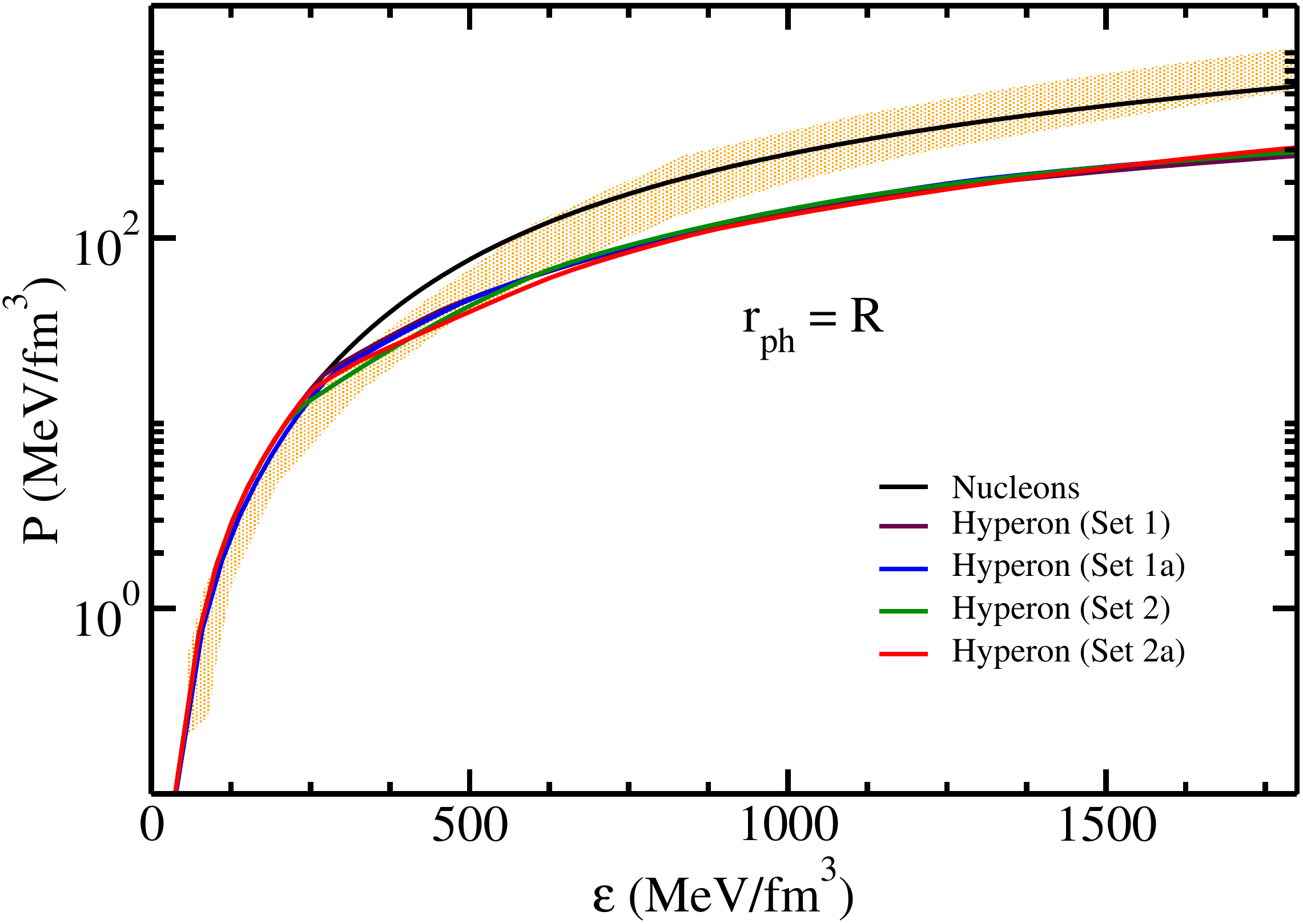}
\caption{\label{fig:2}(Color online) The equation of states obtained for nuclear and hyperon 
matter under charge neutrality as well as the $\beta-$equilibrium condition from G2 force are 
compared with the empirical data (shaded area in the graph) by Steiner et al for $r_{\mathrm ph}$ 
= $R$ with the uncertainty of 2$\sigma$ \cite{stein10}.}
\end{center}
\end{figure}

By imposing the above conditions on the Eqs. (\ref{eqnP},\ref{eqnE}), we have self consistently 
calculated the energy density ${\cal E}$ and pressure density $\cal P$ of hyper-nuclear star 
matter as a function of baryon density. The results obtained for nuclear (i.e., proton and 
neutron) and the octet system (i.e. proton, neutron, $\Lambda^0$, $\Sigma^0$, $\Sigma^+$, 
$\Sigma^-$, $\Xi^0$, and $\Xi^-$) are shown in Fig. \ref{fig:2}. In those curves, there are 
five EoSs for compact star, one from nucleonic matter and four from hyper-nuclear matter for 
various hyperon-meson coupling ratios (given in Table 2). From the figure, it is clearly 
noticed that the EoS from the nucleonic matter is a little stiffer compared to the hyperon 
matter at high density. In other words, the inclusion of hyperons into the compact star, gives 
a softer EoS as compared to the nucleonic matter with respect to the baryon density. Comparison 
with the empirical data for $r_{\mathrm ph}$ = $R$ with the uncertainty 2$\sigma$ of Steiner 
{\it et. al.} \cite{stein02,stein10} (shaded region) is also depicted. Here $R$ and 
$r_{\mathrm ph}$ are for the neutron radius and the photospheric radius, respectively. More 
care inspection to the figure shows that the EoS for nuclear system agrees well with the 
empirical data throughout the densities, but deviates a little for hyperon matter at high 
density. For example, the EoS obtained from the baryon octet family coincides with the 
empirical values up to the density $\sim$ 6$\rho_0$, then it becomes a little softer compared 
to empirical values. Hence, one can interpret that the inclusion of hyperon matter to the 
nucleon system makes the neutron star EoS softer as shown in the figure, which is consistent 
with the other theoretical predictions \cite{ellis91,schaf94,bog81,glend91,lu03,chia09,lu14,sashi07,long12,zhang13,qi15,lim15,biswal16,kras08,oert15,sharma15,haen16,bom16,bend12,mas15,yama14,lona15,drag14,alf07,klahn13,zdun13,chat16,laura17}.

\section{Stellar equations for static and rotating neutron star}
The structure of a spherically symmetric and static compact star, consisting of relativistic 
matter modeled as perfect fluid, can be studied in term of energy density  ${\cal E}$ and the 
pressure density ${\cal P}$ as a function of baryon density using Tolman-Oppenheimer-Volkoff 
equations \cite{tolman39,oppen39}. The general form of the TOV equation is given by:
\begin{equation}
\frac{d{\cal P}}{dr}=-\frac{G}{r}\frac{\left[{\cal E} + \cal P\right ] \left[M+4\pi r^3 
\cal P \right]}{(r-2 GM)},
\end{equation}
\begin{equation}\label{tov2}
\frac{dM}{dr}= 4\pi r^2 \cal E,
\end{equation}
where $G$ and $M (r)$ are the gravitational constant and the enclosed gravitational mass of 
radius $r$, respectively. For a given value of ${\cal P}$ and ${\cal E}$, these equations can 
be integrated from the origin as an initial value problem for a given choice of central energy 
density. The value of $r (= R)$, where the pressure vanishes defines the surface of the star.

\begin{figure}
\begin{center}
\includegraphics[width=0.9\columnwidth]{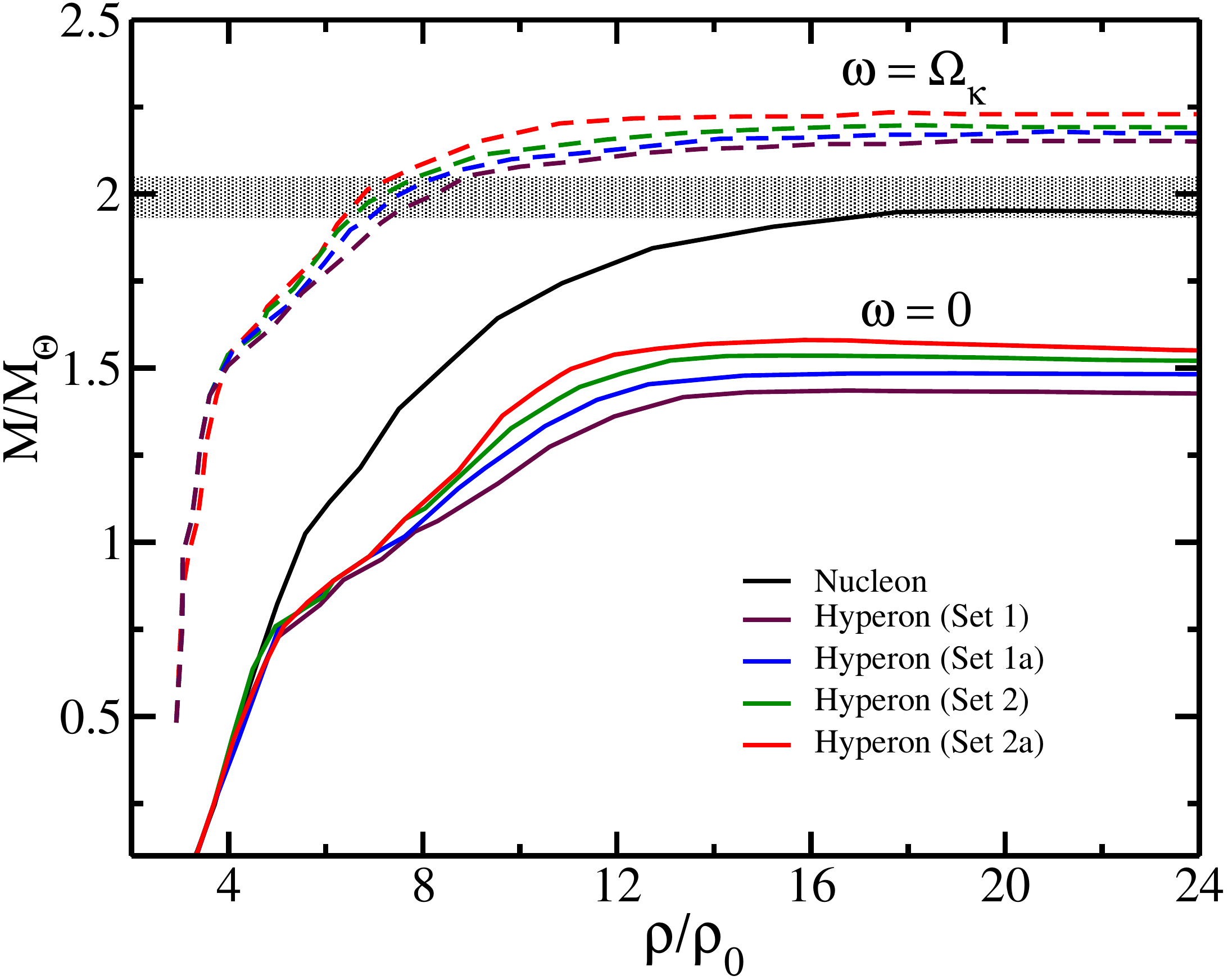}
\caption{\label{fig:3} (Color online) The obtained static (solid line) and rotational (dashed 
line) compact star mass for nuclear and hyper-nuclear matter as a function of baryon density 
are given along with the recent mass observations \cite{demo10,anto13}. See text for details.}
\end{center}
\end{figure}
\begin{figure}
\begin{center}
\includegraphics[width=0.9\columnwidth]{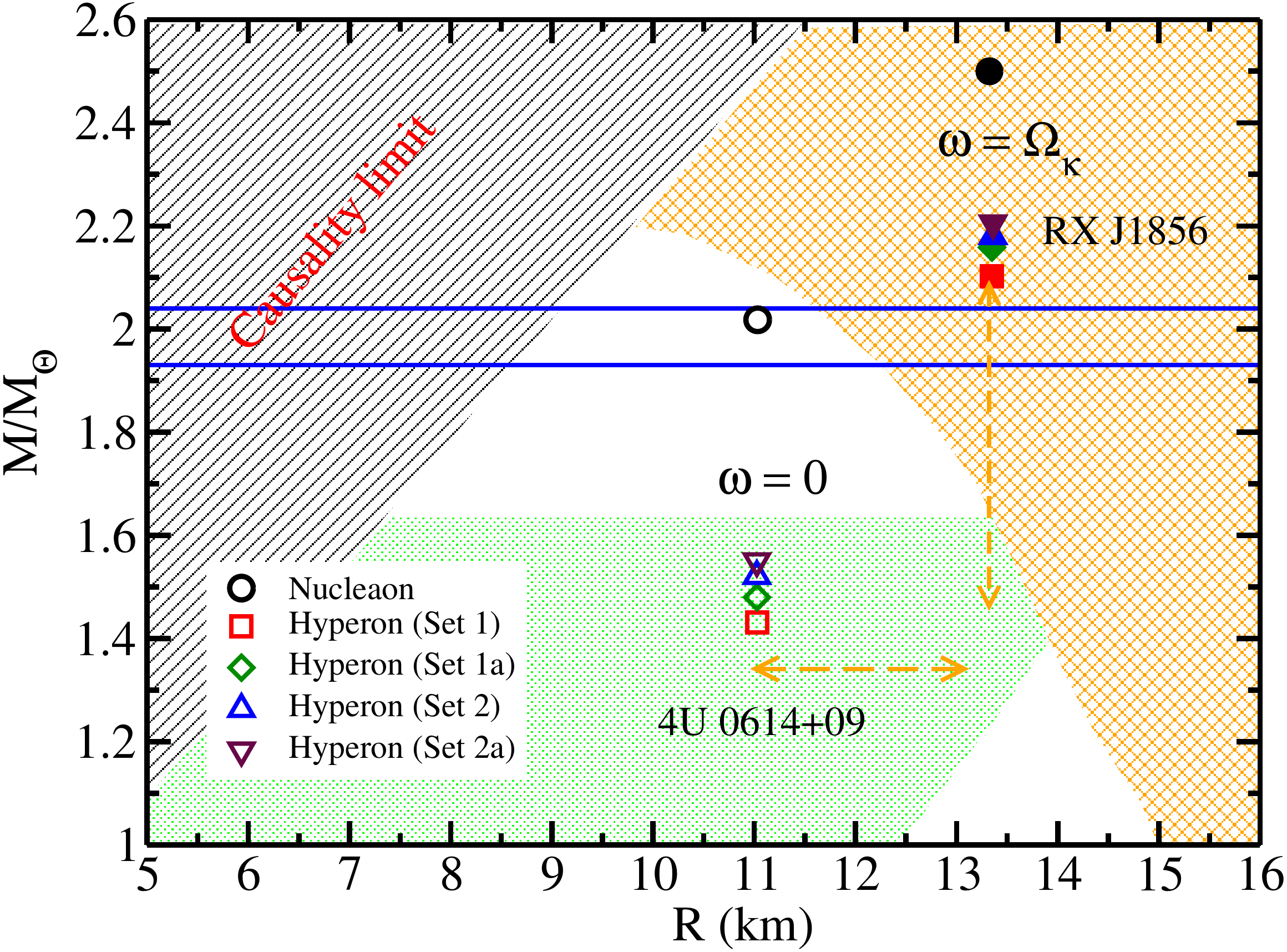}
\caption{\label{fig:4}(Color online) The obtained mass and radius trajectories nuclear and 
hyper-nuclear matter of static and rotational compact stars are compared with the recent mass 
observational datas \cite{demo10,anto13} (the region between two solid horizontal blue line). 
The dashed orange vertical and horizontal lines are stands for the shift of mass and radius 
respectively due to the inclusion of rotational profile to the EoSs of compact star. The 
mass-radius constraints from thermal radiation of isolated neutron star RX J1856 \cite{klahn06} 
(orange hatched area) and from QPOs in the LMXBs 4U 0614+09 \cite{stra00} (green hatched area) 
are given for constraining to the EoSs of compact stars. See text for more details.}
\end{center}
\end{figure}

As we know, the fast rotating relativistic compact stars are sensitive to the stellar mass 
and to the equations of state at high density under $\beta$-equilibrium. Further, the 
rotational instabilities can produce gravitational waves, the detection of which would 
initiate a new field of observational asteroseismology of relativistic stars \cite{satya09}. 
Many interesting phenomena are studied from several independent numerical codes for obtaining 
accurate picture of rotating neutron stars in full general relativistic framework. The metric 
of the space time for a realistic configurations of an axis-symmetric  and stationary rotating 
compact star in spherical polar coordinate is given by \cite{koma89,laar99}:
\begin{eqnarray}
ds^2 &=& -e^{2\nu} dt^{2} + e^{2\alpha}(dr^{2} + r^2 d\theta ^2) \nonumber \\ 
&&+ \mbox{  }e^{2\beta}r^2 \sin^2 \theta(d\phi - \omega dt)^2. 
\end{eqnarray}
Here, ($r$, $\theta$, and $\phi$) are the spherical polar coordinates and the metric potentials 
$\nu$, $\alpha$, $\beta$, $\omega$ are functions of $r$ and $\theta$ only \cite{fried95,laar99}. 
In the limit of perfect fluid, the energy momentum tensor can be given as,
\begin{equation}
T^{\mu \nu} = Pg^{\mu\nu} + ({\cal P + E}) u^{\mu}u^{\nu},
\end{equation}
where $\varepsilon$, $\cal P$, and $u^{\mu}$ are the total energy density, the pressure density 
and the four velocity, respectively \cite{fried95,fried13}. Here, one should follow the three 
basic equations: (i) the Einstein's equation for the metric potentials
\begin{equation}
G_{\mu\nu} = 8\pi T_{\mu\nu}; \nonumber \\
\end{equation}
(ii) the rest mass conservation
\begin{equation}
\bigtriangledown_{\mu} \left( \rho u^{\mu} \right) = 0; \nonumber \\
\end{equation}
and (iii) the stress energy conservation
\begin{equation}
\bigtriangledown_{\nu} T^{\mu\nu} = 0; 
\end{equation}
see Ref. \cite{koma89}, for detailed mathematical expressions. Furthermore the Einstein's 
equation is split into four component for the potentials and the four velocity as,
\begin{equation}
u^{\mu}=\frac{e^{-\nu}}{\sqrt{1-v^2}} (1, 0, 0, \Omega).
\label{u}
\end{equation}
Here, $v$ is the spatial linear velocity with respect to an observer with zero angular  
momentum,
\begin{equation}
v=e^{\beta-\nu} r\; \sin \;\theta (\Omega-\omega).
\end{equation}
Now, we can use the limit on the maximum rotation, by the onset of mass shedding from the 
EoS of the compact star. Here, we have used the geometrized units, $c$ = $G$ = $1$. For 
calculation of the rotational compact star properties like mass, radius and rotational 
frequency, we have used the well known rotational neutron star (RNS) code, which is written 
by Stergioulas $\&$ Friedman \cite{fried95}.

The calculated maximum mass $M$ and the radius $R$ for the static nuclear and hyperon stars 
are obtained from the well-known  TOV equations using the EoSs of E-RMF. The estimated results 
for the maximum mass as a function of density are compared with the observational data from 
pulsars J1614-2230 and J0348+0432 \cite{demo10,anto13}, shown in Fig. \ref{fig:3}. We have 
shown the maximum mass and radius trajectory for the static and rotating compact star in Fig. 
\ref{fig:4}. The dashed orange vertical and horizontal lines in the figure stands for the 
shift of mass and radius respectively due to the inclusion of rotational profile to the EoSs 
of compact star (will discuss in the preceding paragraph). The mass-radius constraints from 
thermal radiation of isolated neutron star RX J1856 \cite{klahn06} (orange hatched area) 
and from QPOs in the LMXBs 4U 0614+09 \cite{stra00} (green hatched area) are given for 
constraining the EOSs of compact stars. Ensuing these recent observations 
\cite{demo10,anto13,klahn06,stra00}, it is clearly illustrated that the maximum 
mass predicted by any theoretical models should reach or near the limit $\sim$ 
2.0$M_\odot$, which is consistent with our present prediction from the EoS of nucleonic 
matter compact star. But, the mass reduced somewhat by inclusion of hyperon matter to the 
EoSs under $\beta$-equilibrium conditions. In other words, the maximum mass obtained from 
nuclear matter EoS is reduced by $\sim$ 0.4$M_\odot$ in presence of hyperon matter core. 
For example, the mass predicted from the nuclear and hyper-nuclear matter are $\sim$ 
2.1$M_\odot$ and $\sim$ 1.5$M_\odot$, respectively for static compact star using the 
TOV equation (see Figs. \ref{fig:3} $\&$ \ref{fig:4}). Hence, one can easily contend that 
the predicted maximum mass from the hyperon matter EoSs is underestimated to the recent mass 
observations \cite{demo10, anto13}, which is the well known {\it Hyperon Puzzle}. It is worth 
mentioning that the results obtained from our calculations are in consistent with the recent 
theoretical predictions \cite{ellis91,schaf94,bog81,glend91,lu03,chia09,lu14,sashi07,long12,zhang13,qi15,lim15,biswal16,kras08,oert15,sharma15,haen16,bom16,bend12,mas15,yama14,lona15,drag14,alf07,klahn13,zdun13,chat16,laura17}. 
Furthermore, there is no significant precise information regarding the constraint on the 
radius $R$ of the compact star. Hence in all cases, the radius of the neutron star corresponding 
to the maximum mass is within the range of $\sim$ 12 km, which satisfy the standard value as 
predicted in the  Ref. \cite{samma09,dutra12} for $214$ Skryme-Hartree-Fock forces.

\begin{figure}
\begin{center}
\includegraphics[width=1.0\columnwidth]{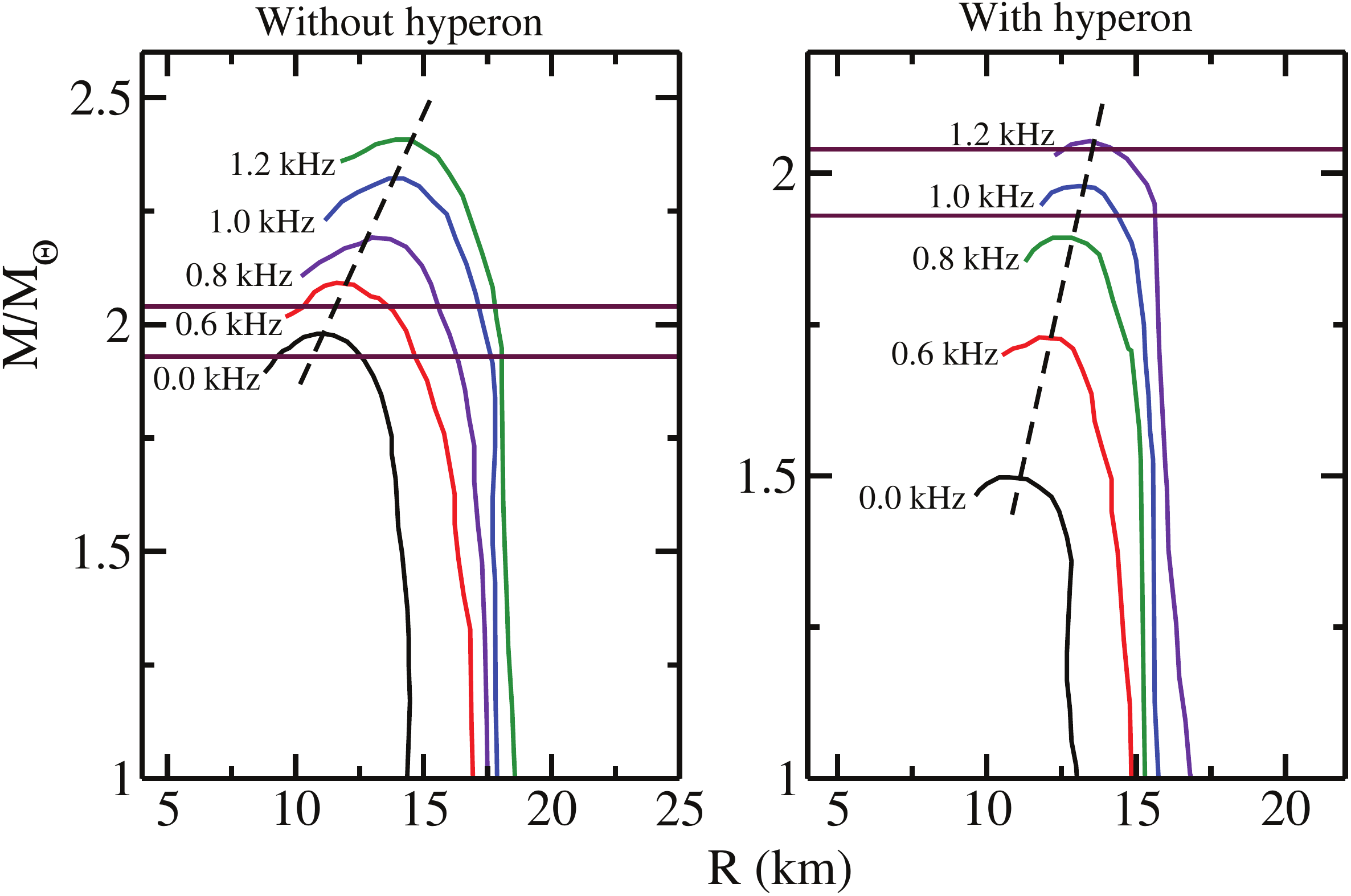}
\caption{\label{fig:5a}(Color online) The gravitational mass $vs.$ radius configuration for 
EoS of nuclear and hyperon matter compact stars with constant rotational frequency. The 
rotational frequency is labeled for each trajectory. The horizontal line in each panel 
represent the recent mass observation \cite{demo10,anto13} are given for a ideal reference 
for the mass and radius. See text for more details.}
\end{center}
\end{figure}

The stationary configurations of a compact star rotating at a given rotational frequency 
$\omega$ forms an one-parameter family characterized by central density $\rho_c$, with 
limitation of two instabilities in the mass $vs.$ equatorial radius plane. For the high 
central density, the axis-symmetric perturbation makes it unstable, as a result the star 
collapse into a Kerr Black hole \cite{kerr63}. In this section we performed the calculations 
for Mass $vs.$ radius relations of very fast rotating (the Keplerian frequency $\omega = 
\Omega_k$) for five different EOSs of compact star using RNS equations. The obtained results 
are given along with the non-rotating compact star properties in Fig. \ref{fig:3} $\&$ 
\ref{fig:4}. From the figures, it is clearly noticed that the rotational profile for the 
EoS at $\omega = \Omega_k$ brings a mass increase by $\sim$ 0.5$M_{\odot}$ and radius by 
$\sim$ 2 km (see Figs. \ref{fig:3} $\&$ \ref{fig:4}). In other words, the maximum 
gravitational mass and radius for a given EoS is enhanced by $\sim$ 0.5$M_{\odot}$ and 
$\sim 2$ km, respectively for a compact star with Keplerian frequency. Quantitatively, the 
mass predicted using EoS of hyperon matter of certain coupling ratio are $\sim 1.5$ (for 
static compact star using TOV equations) and $\sim 2.0$ for stationary rotation (using RNS 
equations). The obtained results for the rotating neutron star properties are consistent 
with the works by Haensel {\it et al.} \cite{haen16}. The crucial question from the 
observational point of view is which the instability with respect to oscillations determine 
the bound state of the compact stars. In order to answer this question, we have calculated 
the gravitational mass-radius trajectories for a broad range of sub-millisecond frequencies, 
0.6$-$1.4 kHz of uniformly rotating compact stars. The obtained results are displayed in 
Fig. \ref{fig:5a} along with the recent mass observation \cite{demo10,anto13}. The left and 
right panel of the figure are corresponding to the EoS of nuclear and hyperon matter compact 
star, respectively. Here, we give the results of mass-shedding for a given rotational 
frequency for only one EoS of hyperon matter obtained from Set 2a \cite{chia09} (see Table 2), 
as a example. We have also found similar results for all other hyperon matter EoSs obtained 
by using the various hyperon-meson couplings listed in Table 2. In the figure, we have only 
shown the mass-shedding points for each frequency of millisecond compact stars with respect 
to axis-symmetric stationary rotation. From the figure, it is clear that the mass and the 
radius increases monotonically with the rotational frequency. Instability with respect to 
the mass-shedding from the equator implies that for a specific rotational frequency the 
gravitational mass and radius should be smaller than the maximum mass at Keplerian limit. 
Hence, one can conclude that the properties of a compact star are influenced with the 
rotational frequency of the compact star but the mass of the hyperon star can only reach 
to the recent mass prediction at Keplerian frequency. Nevertheless, both features i.e. 
hyperon matter in the compact star and the rotation profile are of great practical importance 
and interest. In principle, they are very useful for selecting a perfect EoS to constrain 
the mass measurement and is simultaneously able to work for a wide range of rotational 
frequencies.

\begin{figure}
\begin{center}
\includegraphics[width=1.0\columnwidth]{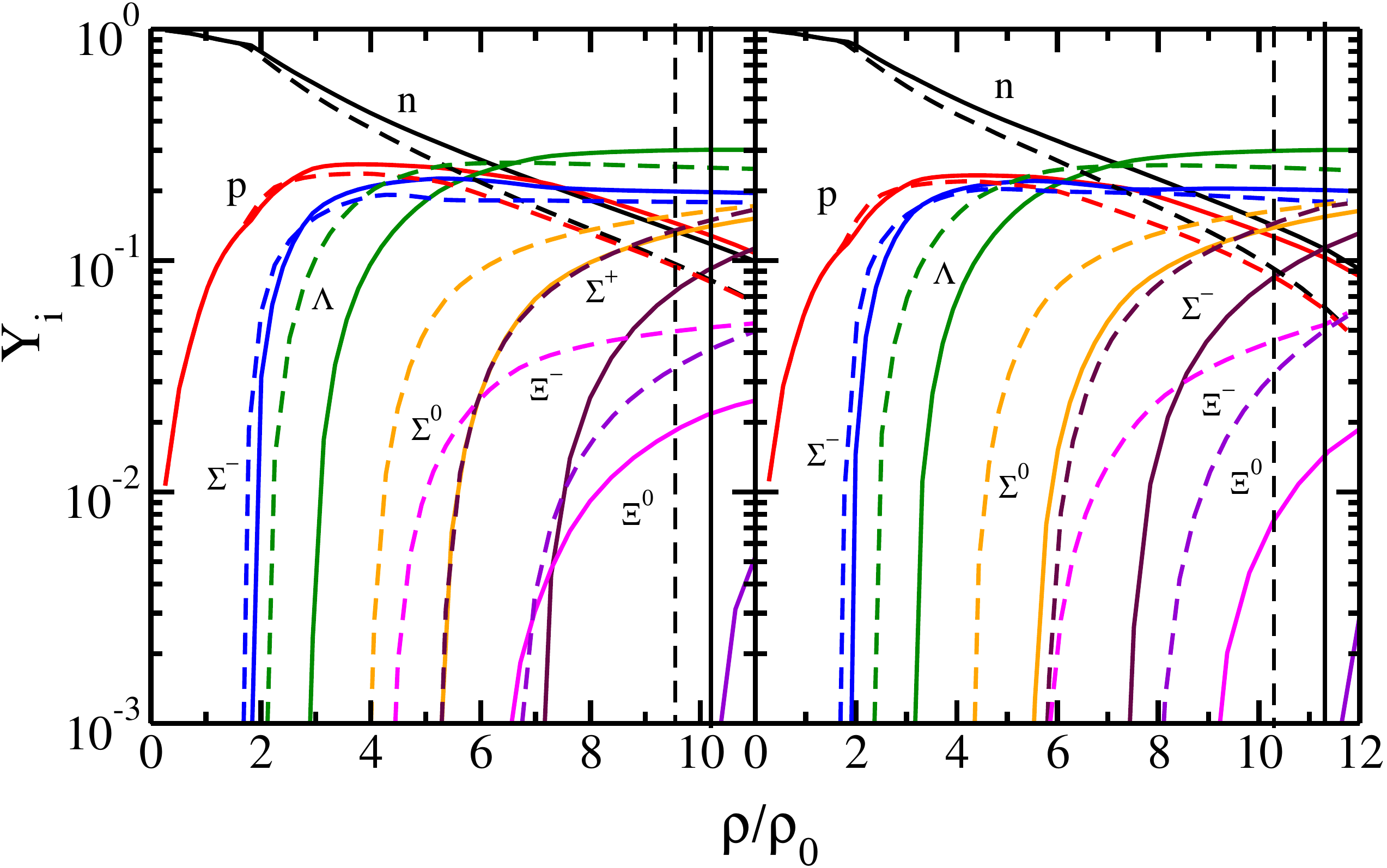}
\caption{\label{fig:5}(Color online) The particle fractions in hyperon matter as a function 
of density. Left panel, for constant coupling ratios \cite{ellis91,schaf94} (Set 1 \& 1a of 
Table 2) and Right panel, for variable coupling ratios \cite{glend91,chia09} (Set 2 \& 2a 
of Table 2). The vertical black lines (solid and dashed) in each panel for the central 
density of the neutron star at the maximum mass. See text for more details.}
\end{center}
\end{figure}

\section{Composition of nuclear matter}
In the above, we have mentioned that there are two different ways to incorporate the 
hyperon-meson coupling ratio into account: (i) constant coupling ratios as assumed by the 
quark model \cite{ellis91,schaf94} (i.e Set 1 \& 1a of Table 2) and (ii) variable couplings 
for different baryons \cite{glend91,chia09} (Set 2 \& 2a of Table 2). The results obtained 
for the yield as a function of density using these coupling ratios are shown in Fig. 
\ref{fig:5}. The left and right panel of the figure represents the prediction from the 
constant and variable couplings, respectively. Furthermore, the solid and dashed lines 
in each panel stand for the two sets of coupling for a specific approach in the 
hyperon-meson interactions. From the figure, it is easily noticed that the $\Sigma^-$ is 
generated at $\rho_B \sim$ 2.0$\rho_0$ and its fraction increases rapidly to a saturation 
at $\rho_B$ $\sim$ 3.1$\rho_0$. Similarly, the $\Lambda$-hyperon generated at 
$\sim$2.1$\rho_0$ and the yield becomes constant at $\sim$5.1$\rho_0$. From the figure, 
one can clearly notice that the yields for the two coupling ratios for each (constant 
\& variable) approach have almost similar predictions. A careful investigation between 
two methodologies are found to differ a little from each other for the yields with respect 
to density. In other words, the procreation of hyperons as a function of density are 
altered a little for two approaches. Hence, one can conclude that the yields are 
marginally dependent on coupling ratios.

\section{Summary and Conclusions}
In the present study, we  applied an effective field theory motivated relativistic mean 
field approach to the nuclear matter, where we investigated the influence of hyperon matter 
and the rotational profile on the properties of the compact star. We have generated four EoSs 
for the hyperon matter stars using various hyperon-meson coupling ratios. All the predicted 
EoSs for the hyperon stars have produced a maximum mass $\sim$ 1.5$M_{\odot}$, which is lower 
compared to the recent star observation data \cite{demo10,anto13}. The maximum radii 
corresponding to the masses are located in the range of 11$-$13 km. The influence of hyperon 
in the present study are consistent with the predictions of other theoretical calculations
\cite{ellis91,schaf94,bog81,glend91,lu03,chia09,lu14,sashi07,kras08,oert15,sharma15,haen16,bom16,bend12,mas15,yama14,lona15,drag14,alf07,klahn13,zdun13,chat16,laura17}, both early and the 
recent ones in their mass spectrum under charge neutrality and $\beta$-equilibrium conditions. 
In continuation to the {it Hyperon Puzzle}, we have included the rotational profile under 
axis-symmetric constant rotation with Keplerian frequency into account using rotational star 
equation, which increase the mass of the compact star by $\sim 0.5$ solar mass in magnitude 
as compared to the static case. In other words, the EoSs of hyperon matter along with the 
rotational profile at Keplerian limit predict the extreme mass $\sim$ 2.1$M_\odot$, which 
are in good agreement with the late observation \cite{demo10,anto13}. Furthermore, we have 
also calculated the mass-shedding for a given EoS for rotational frequency sequences of 
uniformly rotating compact stars. We found that the mass and the radius are increased 
monotonically with the rotational frequency for millisecond pulsar. Instability with respect 
to the mass-shedding from the equator implies that for a specific rotational frequency the 
gravitational mass and radius should be smaller than the maximum mass at Keplerian limit. 
From the above analysis, one can conclude that, the properties of a compact star are 
influenced with the rotational frequency but the mass of the hyperon star can only reach 
to the recent mass prediction at Keplerian frequency. In other words, the inclusion of 
rotational profiles to the neutron stars EoSs are not a full phase solution for the 
{\it Hyperon Puzzle}. Hence, it is important to consider other features along with the 
rotational profile in predictions of the hyperon star properties. In principle, the 
constituent of the core of the compact star is very useful for selecting a perfect EoS
to constrain the mass measurement and is simultaneously able to work for a wide range 
of rotational frequencies. In summary, the present study is useful in the admiration of 
the expectation of static and stationary rotating hyper-nuclear compact stars at high 
density.

\section*{Acknowledgments}
This work has been supported by the 973 Program of China (Grant No. 2013CB834400), the 
NSF of China (Grants No. 11120101005, No. 11275248, and No. 11525524), the Chinese Academy 
of Sciences (Grant No. KJCX2-EW-N01) and the FAPESP Project No. 2014/26195-5 and and by 
the CNPq - Brasil. 

\end{document}